\documentclass[lettersize,journal]{IEEEtran}
\usepackage{amsmath,amsfonts}
\usepackage{algorithm}
\usepackage{algpseudocode}
\usepackage{array}
\usepackage[caption=false,font=normalsize,labelfont=sf,textfont=sf]{subfig}
\usepackage{textcomp}
\usepackage{stfloats}
\usepackage{url}
\usepackage{verbatim}
\usepackage{graphicx}
\usepackage{cite}
\usepackage{xcolor}
\usepackage{amsthm}
\usepackage{amssymb}
\usepackage{amsmath,lipsum}
\usepackage{epstopdf}
\usepackage{bm}
\usepackage{makecell}
\usepackage{cuted}

\begin{document}

\title{\huge DSRDM: Digital Signal Recovery Diffusion Model for Semantic Communications}
\vspace{-4mm}
\author{Zhigang Yan and Dong Li,~\IEEEmembership{Senior Member, IEEE }
\vspace{-8mm}
\thanks{Zhigang Yan and Dong Li are with the School of Computer Science and Engineering, Macau University of Science and Technology, Macau, China. (e-mails: 3220005784@student.must.edu.mo; dli@must.edu.mo). (\textit{Corresponding Author: Dong Li}.)
}
}



\maketitle
\pagestyle{empty}
\thispagestyle{empty}

\begin{abstract}
Diffusion model (DM) has recently appeared as a promising type of generative model for AI-generated content, which has been widely used for image reconstruction, generation, and channel denoising in semantic communication (SemCom) due to its strong generation capacity. However, most of existing works regarding SemCom remain confined to the image or text transmission, and neglect the commonly adopted digital signals in wireless systems. In this letter, in order to address this gap, we propose and investigate a digital signal recovery diffusion model (DSRDM) for SemCom. Specifically, DSRDM encodes digital signals by gradually adding Gaussian signals to images in the forward diffusion process of DM. After the encoded Gaussian signals embedded in the carrier image are sent to the receiver, it recovers the digital signals by predicting the added Gaussian signals iteratively in the reverse diffusion process. Moreover, to reduce the computation complexity of DSRDM, a signal adding approach is designed to avoid the retraining latency. In particular, we use the latent representation of images instead of themselves as the carrier for digital signals in DSRDM to reduce the inference latency. 
\end{abstract}
\vspace{-1mm}
\begin{IEEEkeywords}
Semantic communication, diffusion model.
\end{IEEEkeywords}
\vspace{-3mm}
\section{Introduction}
\IEEEPARstart{D}{iffusion} models (DM) have recently attracted significant attention in AI-generated content \cite{diffusion}. Due to the strong feature extraction and generative capabilities, DM has been widely applied to semantic communication (SemCom) for images, audios and videos transmission \cite{image1,image2,audio,video}. On the other hand, SemCom aims to transmit semantic information rather than just raw bits, making it more robust to noise and more efficient in the bandwidth usage. The semantic information of the source data are encoded by deep learning (DL) models, such as convolutional neural network, Transformer and DM \cite{video,text,dm}. Since DL models have been widely recognized to be able to extract features from structured data, such as images, text, and videos \cite{dl}. However, unlike other DL models that are usually only used as encoders and decoders in SemCom \cite{semcom}, DMs are also used in SemCom systems as channel denoising module \cite{cddm,scdm} and channel enhancer \cite{dmce} based on its powerful denoising and generation capabilities. Specifically, channel denoising DM \cite{cddm} and score-based channel denoising DM \cite{scdm} are modules for removing channel noise and purifying received signals. DM channel enhancer \cite{dmce} is designed to suppress the noise in the channel state information (CSI) estimation by learning the distribution of data and received signals. All these DM-based modules focus on the denoising ability of DM. 

However, in SemCom systems, little attention has been drawn to digital signals since they do not have obvious semantic information. Moreover, large-scale transmission of digital signals is necessary in some scenarios, such as uploading tabular data collected by sensors and updating model parameters in wireless federated learning \cite{fl}. There has been limited effort to solve the problem of digital signals transmission via SemCom, and \cite{mae} was the first attempt to this problem to the best of our knowledge, which considered using mask images by multiplying with digital signals, and then these masked images were encoded as latent representations by Masked Autoencoders (MAE) and transmitted to receivers. Besides, the Model Shift Modulation (MSM) was proposed for SemCom in \cite{msm}, which mapped the digital signal to distinct shifts in the image’s feature map for transmitting it over the image transmission. However, the size of the image/feature map in \cite{mae} and \cite{msm} limit the number of bits that can be transmitted in a single transmission.

In order to solve the problem of constrained digital bits delivery in existing works, in this letter, we propose a digital signal recovery DM (DSRDM) as the encoder and decoder for SemCom to transmit the digital signals. Compared to previous encoders, DM shows a greater potential in encoding digital signals since the forward diffusion process of DM allows more information to be added to an image. Specifically, DSRDM adds Gaussian signals to images across multiple steps of the forward process, instead of adding Gaussian noise. Then the noisy images as carriers which carry a series of Gaussian signals are transmitted to the receiver. Finally, the receiver uses a DM to iteratively predict the added Gaussian signals and reconstruct the digital signal on the reverse diffusion process. The main contributions are summarized as follows:
\begin{itemize}
\item \textbf{A novel transmission approach for digital signals via DM.} We propose a novel DSRDM encoder and decoder for digital signal transmission in SemCom. By grouping the original bits into complex constellation symbols with $M$-ary modulation, the original digital signal is converted into an approximate Gaussian signal, which is then injected into the forward diffusion process. The added signal can be predicted and recovered by the denoising capability of a well-trained DM.
\item \textbf{A training costs minimization algorithm for DSRDM.} To reduce the training costs of DSRDM, we design a training-free algorithm for DSRDM. It maps the added signals to the known statistical distribution of the noise expectation of the pre-trained DMs, which makes DSRDM can use pre-trained models without retraining. 
\item \textbf{A latency-efficient design for DSRDM for fast inference.} To avoid the large inference delay of DM, we encode the latent representation from images before transmission, and use it instead of the image itself as the carrier for digital signals in DSRDM to reduce the complexity of inference at the receiver.
\end{itemize}

\section{System Model of SemCom}

In this section, we introduce the SemCom system with the proposed DSRDM. Specifically, the source data $\mathbf{S}$ are encoded as its semantic information $\mathbf{Z}$ by the semantic encoder. Then, they are encoded into the complex-valued signal $\mathbf{X}$ by the channel encoder, which can be expressed as
\vspace{-1mm}
\begin{equation}
	\label{encode}
	\mathbf{X} = f_{\rm ce}\big(f_{\rm se}(\mathbf{S})\big),
	\vspace{-1mm}
\end{equation}
where $\mathbf{X}\in\mathbb{C}^{m}$ and $m$ denotes the number of channel used. $f_{\rm se}(\cdot)$ and $f_{\rm ce}(\cdot)$ are the semantic and channel encoding function, respectively. In addition, $\mathbf{Z}=f_{\rm se}(\mathbf{S})\in\mathbb{R}^{2m}$. Let $x_{i}$ and $z_{i}$ be the elements of $\mathbf{X}$ and $\mathbf{Z}$, we have $x_{i}=z_{i}+jz_{i+m}$, for $i=1,\cdots,m$, where $j=\sqrt{-1}$ is the imaginary unit. Then $\mathbf{X}$ is sent to the receiver over the wireless channel, and the received signal is
\begin{equation}
	\label{trans}
	\mathbf{Y} = \mathbf{H}\mathbf{X}+\mathbf{N},
	\vspace{-1mm}
\end{equation}
where $\mathbf{H}=\mathrm{diag}(h_{i})\in\mathbb{C}^{m\times m}$ is the channel matrix and $\mathbf{N}\sim\mathcal{CN}(\mathbf{0},2\sigma_{n}^{2}\mathbf{I})\in\mathbb{C}^{m}$ is the additive white Gaussian noise (AWGN). Besides, considering the
effects of Rician fading, we have $h_{i}=\sqrt{\frac{k}{k+1}}+\sqrt{\frac{1}{k+1}}h_{r,i}$ where $k$ denotes the ratio of the direct and non-direct radio wave power and $h_{r,i}$ are independent and identically distributed (i.i.d.) Rayleigh fading gains for all $i$. 

Next, the receiver decodes $\mathbf{Y}$. If the channel decoding function is $f_{\rm cd}(\cdot)$, the received semantic information vector is given by $\mathbf{\hat{Z}} = f_{\rm cd}(\mathbf{Y})=[\mathrm{Re}(\mathbf{\hat{X}}),\mathrm{Im}(\mathbf{\hat{X}})]$, where $\mathrm{Re}(\cdot)$ and $\mathrm{Im}(\cdot)$ the real and imaginary parts of the complex vector.  We assume that the CSI is perfect, so we have $\mathbf{\hat{X}}=\mathbf{\hat{H}}\mathbf{Y}=\mathbf{X}+\mathbf{\hat{H}}\mathbf{N}$, where $\mathbf{\hat{H}}=[(\mathbf{H})^{H}\mathbf{H}]^{-1}(\mathbf{H})^{H}$. 
Accordingly, $\mathbf{\hat{Z}}$ can be rewritten as
\begin{equation}
	\label{decode1}
	\mathbf{\hat{Z}} = \mathbf{Z} + \mathbf{\hat{N}},
	\vspace{-1mm}
\end{equation}
where $\mathbf{\hat{N}}=[\mathrm{Re}(\mathbf{\hat{H}}\mathbf{N}),\mathrm{Im}(\mathbf{\hat{H}}\mathbf{N})]\sim\mathcal{N}(\mathbf{0},\sigma_{n}^{2}[(\mathbf{H})^{H}\mathbf{H}]^{-1})$. Finally, $\mathbf{\hat{Z}}$ is input into the semantic decoder to recover the received data $\mathbf{\hat{S}}$, which can be given by 
\begin{equation}
	\label{decode2}
	\mathbf{\hat{S}} = f_{\rm sd}(\mathbf{\hat{Z}}),
	\vspace{-1mm}
\end{equation}
where $f_{\rm sd}(\cdot)$ denotes the semantic decoding function. 

The semantic encoder and decoder are typically trained jointly in an offline setting using a large dataset, often referred to as shared knowledge. The primary objective during training is to minimize the discrepancy between $\mathbf{S}$ and $\mathbf{\hat{S}}$, thereby ensuring that the semantic information encoded by the transmitter is accurately recovered by the receiver.

\section{Proposed DSRDM}

DM was originally proposed as a generative model, that learns to reverse the process of gradual noise-adding, enabling it to generate complex data distributions. Specifically, in previous studies on DM, finishing such generation tasks was proved to be equivalent to minimizing the mean square error (MSE) of the noise predicted in the reverse process, and the true noise added in the forward process, so its loss function is
\begin{equation}
	\label{loss}
	\mathcal{L}_{\mathrm{DM}}\triangleq\mathbb{E}_{t \sim [1, T], \mathbf{x}_t, \boldsymbol{\epsilon}_t} \big[\|\boldsymbol{\epsilon}_t - \boldsymbol{\epsilon}_\theta(\mathbf{x}_t, t)\|^2 \big].
\end{equation}
where $\boldsymbol{\epsilon}_{\theta}(\mathbf{x}_{t},t)$ is the predicted noise in the reverse process and $\boldsymbol{\epsilon}_{t}$ is the true noise be added in the forward process, which is given by  
\begin{equation}
	\label{forward}
	\mathbf{x}_{T}=\sqrt{\bar{\alpha}_{T}}\mathbf{x}_{0}+\sqrt{1-\bar{\alpha}_{T}}\boldsymbol{\epsilon},
\end{equation}
where $\mathbf{x}_{0}$ is the real data, $\bar{\alpha}_{T}=\prod_{t=1}^{T}\alpha_{t}$, and $\alpha_{t}\in(0,1)$ is the hyperparameter of DM. 
\begin{figure}[!t]
	\centering
	\includegraphics[width=2.5in]{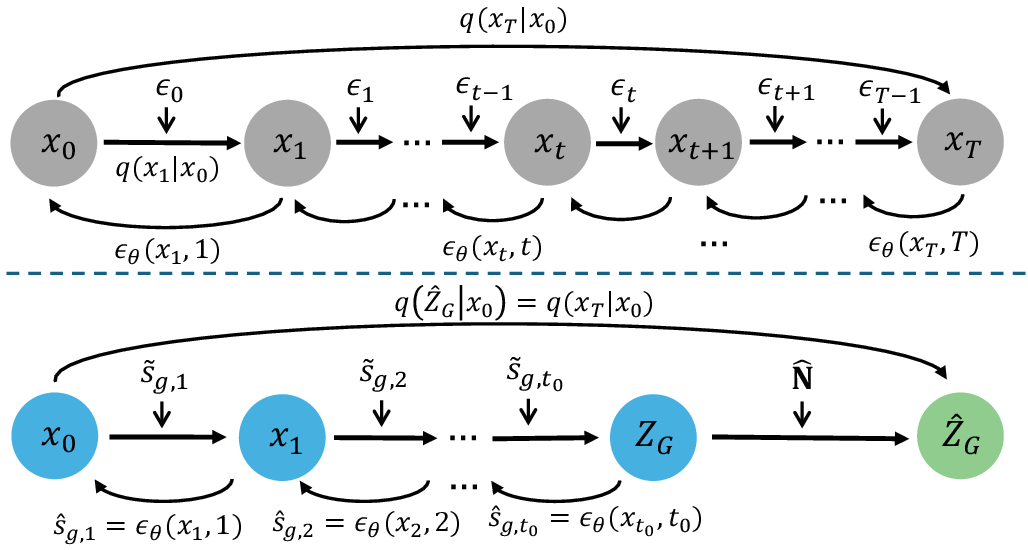}
	\caption{A schematic diagram of proposed DSRDM.}
	\label{fig1}
\end{figure}
This loss function shows that the well-trained DM has excellent performance in Gaussian noise prediction. From \eqref{loss}, the added Gaussian noise added in the forward diffusion process can be accurately predicted by a well-trained DM in its reverse diffusion process. Motivated by this, a novel SemCom system for Gaussian signal transmission based on the strong prediction capacity of DM, DSRDM is proposed. The entire digital signal transmission process based on DSRDM is summarized as follows:
\begin{itemize}
	\item Discrete digital source sequence are grouped to complex constellation symbols by an $M$-ary modulation. When $M$ is large enough, the original digital signals are converted into approximately Gaussian-distributed signals.\footnote{Although the modulated signals are not Gaussian when $M$ is small, we also verify that it could be predicted well by DSRDM in simulations. It is verified that a trained DM can well predict signal with different $M$ without retraining.}
 	\item Encoded Gaussian signals are reshaped in the same size of the carrier images. They are then embedded into this carrier image across multiple steps of the forward diffusion process.
	\item At the transmitter side, instead of sending the structured data itself, the goal is to transmit the Gaussian signals embedded within the diffusion process. The resulting noisy samples are transmitted to the receiver over a communication channel.
	\item At the receiver side, the trained DM is used to predict the added Gaussian noise at each step. Since DM was trained using the MSE loss between true and predicted noise, it retains a strong capability for accurate noise estimation, even if the underlying goal is not image reconstruction. After the DM predicts the noise at each step, the inverse transformation is applied to recover the Gaussian signals.
	\item Finally, the recovered signals are demodulated back to bitstream. Specifically, each received point is then compared with all ideal constellation points, and the nearest one is selected. Each ideal constellation point is mapped back to its associated binary sequence, reconstructing the transmitted bits.
\end{itemize}

This framework allows the DM to act as a signal predictor rather than a generative model, enabling unstructured digital signal transmission by SemCom. A schematic diagram of DSRDM is shown in Fig. 1. More specific details about the proposed DSRDM are introduced in following subsections.\footnote{In the following introduction, we will skip the steps of Gaussian approximation modulation and demodulation of the signal, and focus on the encoding and recovery of Gaussian signals. Performance evaluations of the entire digital signal transmission process will be provided in the simulation results.}

\subsection{Signal Recovery Diffusion Model}

As shown in \eqref{loss} and \eqref{forward}, the outputs of DM are the predicted Gaussian signal $\boldsymbol{\epsilon}_{t}\sim\mathcal{N}(\mathbf{0},\mathbf{I})$ from $t=1$ to $T$, which are added to $\mathbf{x}_{0}$ in the forward diffusion process. Thus, in the proposed SemCom system, the transmitter encodes the source Gaussian data $\mathbf{S}\sim\mathcal{N}(\mu_{s},\sigma_{s}^{2}\mathbf{I})\in\mathbb{R}^{L}$, where $L$ is the length of $\mathbf{S}$ into the standard Gaussian distribution by $\mathbf{s}_{g}=\sigma_{s}^{-1}(\mathbf{S}-\mu_{s})$, where $\mathbf{s}_{g}$ is the signal with standard Gaussian distribution. Then, DSRDM further encodes $\mathbf{s}_{g}$ into the semantic information vector $\mathbf{Z}_{g}$ by replacing $\boldsymbol{\epsilon}_{t}$ in the forward diffusion process with $\mathbf{s}_{g}$, and adds it to a carrier image $\mathbf{x}_{0}\in\mathbb{R}^{H\times H\times 3}$, where $H\times H$ is the size of the image, respectively. Based on \eqref{forward}, this encoding operation is given by
\begin{equation}
	\label{srdm1}
	\mathbf{Z}_{G}=\sqrt{\alpha_{T}}\mathbf{x}_{T{-}1}+\sqrt{1{-}\alpha_{T}}\mathbf{\tilde{s}}_{g}=\sqrt{\bar{\alpha}_{T}}\mathbf{x}_{0}+\sqrt{1{-}\bar{\alpha}_{T}}\mathbf{\tilde{s}},
\vspace{-1mm}
\end{equation}
where $\mathbf{\tilde{s}}_{g}$ is $\mathbf{s}_{g}$ reshaped into $\mathbb{R}^{H\times H\times 3}$ and $\mathbf{\tilde{s}}$ merges all $T$ Gaussian signals $\mathbf{\tilde{s}}_{g}$.

Then, $\mathbf{Z}_{G}$ is sent to the receiver over the wireless channel. According to \eqref{decode1}, after channel decoding, the received $\mathbf{Z}_{G}$ (e.g., $\mathbf{\hat{Z}}_{G}$) is $\mathbf{\hat{Z}}_{G}=\mathbf{Z}_{G}+\mathbf{\tilde{N}}$, where $\mathbf{\tilde{N}}\in\mathbb{R}^{H\times H\times 3}$. Furthermore, based on \eqref{srdm1}, the input of semantic decoder at the receiver can be written as
\begin{equation}
	\vspace{-1mm}
		\label{srdm2}
		\mathbf{\hat{Z}}_{G}=\underbrace{\sqrt{\bar{\alpha}_{T}}\mathbf{x}_{0}}_{\mathbf{x}_{0}'}+\underbrace{\sqrt{1-\bar{\alpha}_{T}}\mathbf{\tilde{s}}_{g}+\mathbf{\tilde{N}}}_{\boldsymbol{\epsilon}_{\rm mix}}.
\vspace{-1mm}
\end{equation}
Thus, $q(\mathbf{\hat{Z}}_{G}|\mathbf{x}_{0})\sim\mathcal{N}(\mathbf{x}_{0}',\mathbf{W})$, where $\mathbf{W}\triangleq(1-\bar{\alpha}_{T})\mathbf{I}+\sigma_{n}^{2}[(\mathbf{H})^{H}\mathbf{H}]^{-1}=\mathrm{diag}(1-\bar{\alpha}_{T}+\sigma_{n}^{2}h_{i}^{-2})$. By comparing \eqref{forward} and \eqref{srdm2}, we can find that if the $\boldsymbol{\epsilon}_{\rm mix}$ is large enough, such as a larger $L$ or $\sigma_{n}^{2}$, to make $\mathbf{\hat{Z}}_{G}\approx \mathbf{x}_{T}$, DSRDM can be trained on $\mathbf{x}_{T}$ instead of $\mathbf{\hat{Z}}_{G}$ to remove the channel noise and recover the source data $\mathbf{S}$ by predicting $\boldsymbol{\epsilon}_{\rm mix}$.

Specifically, if we let the carrier image of DSRDM is $\mathbf{x}_{0}$, similar to \eqref{srdm2} with $\bar{\alpha}_{T}$, after $T$ steps of the forward process of DSRDM, we define
\begin{equation}
	\vspace{-1mm}
		\label{train1}
		\mathbf{x}_{T}=\sqrt{\bar{\alpha}_{T}}\mathbf{x}_{0}+\sqrt{1-\bar{\alpha}_{T}}\mathbf{W}'\boldsymbol{\epsilon},
	\vspace{-1mm}
\end{equation}
where $\mathbf{W}'=(1-\bar{\alpha}_{T})^{-\frac{1}{2}}\mathbf{W}$. Thus, by comparing \eqref{srdm2} and \eqref{train1}, it is easy to check that the Kullback-Leibler (KL) divergence between $q(\mathbf{\hat{Z}}_{G}|\mathbf{x}_{0})$ and $q(\mathbf{x}_{T}|\mathbf{x}_{0})$ is $0$. It means that we can generate $\mathbf{x}_{T}$, which follows the same distribution as $\mathbf{\hat{Z}}_{G}$. Thus, the proposed DSRDM can be trained on the generated $\mathbf{x}_{T}$ instead of the real received $\mathbf{\hat{Z}}_{G}$ by minimizing \eqref{loss}. Then, at the receiver, the trained DSRDM is used to obtain the $\boldsymbol{\epsilon}_\theta(\mathbf{x}_t, t)$ as the recovered $\mathbf{\tilde{s}}_{g}$ added in the $t$-th step of the forward process.\footnote{The above Gaussian signal encoding scheme of DSRDM (i.e., \eqref{srdm1}) can be extended to the multi-user case, where multiple users' bits can be either transmitted via different images or different steps in the diffusion space for one single image.}

\vspace{-2mm}
\subsection{Training-free Algorithm for DSRDM}

Unlike traditional DM for generation tasks which requires the specific datasets to train, DSRDM only need to predict the added Gaussian signal and the structured carrier image $\mathbf{x}_{0}$ is just a medium rather than the primary data of interest. All DMs are naturally trained to do it whatever the training data is, and the pre-trained DM has already learned a strong prediction prior, meaning it can effectively track and predict Gaussian signal added in $\mathbf{x}_{0}$. Therefore, if the added Gaussian-modulated signals follow the same noise statistics to what the pre-trained model was trained on, the pre-trained DM can be applied on DSRDM directly to eliminate the training costs, without retraining or fine-tuning.

To this end, we map the added Gaussian signals of DSRDM into the same statistical distribution that the pre-trained DM expects. Specifically, we assume that the noise adding algorithm of the pre-trained DM is 
\begin{equation}
	\vspace{-1mm}
	\label{pre}
	\mathbf{x}_{T}^{(p)}=\sqrt{\bar{\alpha}_{T}^{(p)}}\mathbf{x}_{0}+\sqrt{1-\bar{\alpha}_{T}^{(p)}}\boldsymbol{\epsilon},
	\vspace{-1mm}
\end{equation}
where $\bar{\alpha}_{T}^{(p)}$ is known. Thus, when we add $\mathbf{\tilde{s}}_{g}$ by the following forward diffusion process, which is
\begin{equation}
		\label{pre2}
		\mathbf{x}_{T}^{(s)}=\boldsymbol{\bar{\alpha}}_{T}\mathbf{x}_{0}+\boldsymbol{\bar{\beta}}_{T}\mathbf{\tilde{s}}_{g},
	\vspace{-1mm}
\end{equation}
where $(\boldsymbol{\bar{\alpha}}_{T},\boldsymbol{\bar{\beta}}_{T})=\big(\mathrm{diag}\big[\sqrt{\bar{\alpha}_{i,T}}\big],\mathrm{diag}\big[\sqrt{1-\bar{\alpha}_{i,T}}\big]\big),i=1,\cdots,H$. Then, let $\bar{\alpha}_{i,T}=\bar{\alpha}_{T}^{(p)}-\sigma_{n}^{2}h_{i}^{-2}$ and substitute it into \eqref{pre2} and \eqref{srdm2}, after transmitting over wireless channel, the result of the forward process of DSRDM (e.g., $\mathbf{x}_{T}^{(s)}$) is expressed as
\begin{equation}
	\begin{aligned}
		\label{pre4}
		\mathbf{\hat{Z}}_{G}=\mathbf{x}_{T}^{(s)}+\mathbf{\tilde{N}}=\sqrt{\bar{\alpha}_{T}^{(p)}}\mathbf{x}_{0}^{(s)}+\sqrt{1-\bar{\alpha}_{T}^{(p)}}\boldsymbol{\epsilon},
	\end{aligned}
\end{equation}
where $\mathbf{x}_{0}^{(s)}=\mathrm{diag}\big[\sqrt{\bar{\alpha}_{T}^{(p)}/\bar{\alpha}_{i,T}}~\big]\mathbf{x}_{0}$. In addition, since the goal of DSRDM is accurate signal transmission rather than image reconstruction, replacing $\mathbf{x}_{0}$ with $\mathbf{x}_{0}^{(s)}$ does not affect the application of pre-trained models to DSRDM. Furthermore, since the KL divergence between $q(\mathbf{\hat{Z}}_{G}|\mathbf{x}_{0}^{(s)})$ and $q(\mathbf{x}_{T}^{(p)}|\mathbf{x}_{0}^{(s)})$ is $0$ when $\bar{\alpha}_{i,T}{=}\bar{\alpha}_{T}^{(p)}{-}\sigma_{n}^{2}h_{i}^{-2}$, the pre-trained DM can predict the added Gaussian signals on $\mathbf{x}_{0}^{(s)}$ from received $\mathbf{\hat{Z}}_{G}$ directly. 

In summary, this training-free algorithm for DSRDM transforms the Gaussian signals to match the known statistics of the noise expectation of the pre-trained DM, enabling direct use of the model without retraining. Since DM is typically trained using Gaussian noise and the added noise at each step follows a known distribution, transforming the information-bearing Gaussian signals to conform to this same distribution, pre-trained DM can interpret them as standard diffusion noise. This allows the pre-trained model to accurately estimate these signals during the reverse process since it is originally optimized using the MSE loss to predict such noise. 
\subsection{Latency Efficient Designs for DSRDM}

In addition to the training costs, decoding latency is also a key concern for DM-based communication systems, since the decoding process of DM is a step-by-step reverse process. Thus, the latency caused by model inference is a challenge for DSRDM application in digital signal transmission. 

To address this issue, we replace the traditional DM in DSRDM with Latent Diffusion Model (LDM),  which essentially means embedding signals into and decoding from the latent space rather than the image space, so the carrier for signal transmission will be the latent representation of image instead of the image itself. Compared with the original image, its latent representation has a smaller spatial size, which means its fewer parameters and faster inference at the receiver. 

\begin{figure}[!t]
	\centering
	\includegraphics[width=2.6in]{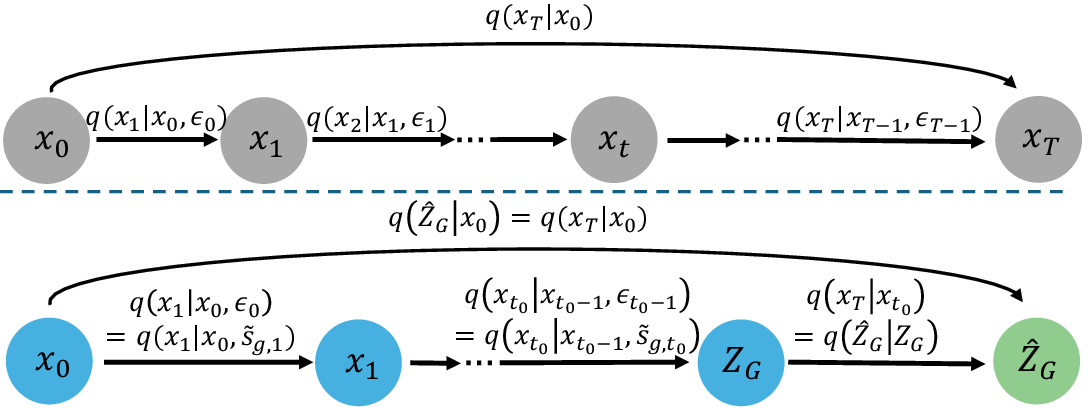}
	\caption{The entire digital signal transmission process based on DSRDM.}
	\label{fig2}
\end{figure}

The changes with LDM replacing traditional DM in DSRDM includes \textit{a) Carrier domain switch} and \textit{b) Modified signal embedding}. Specifically, the carrier of DSRDM with traditional DM is the image (e.g., $\mathbf{x}_{0}\in\mathbb{R}^{H\times W\times 3}$), but LDM operates in a compressed latent space. The carrier is the latent representation of the original image encoded by a learned encoder. Thus, similar to \eqref{pre2}, the signal adding scheme in the forward process of DSRDM with LDM is
\begin{equation}
	\begin{aligned}
		\label{ldm}
		\mathbf{z}_{T}=\mathrm{diag}\big[\sqrt{\bar{\alpha}_{i,T}}\big]\mathbf{z}_{0}+\mathrm{diag}\big[\sqrt{1-\bar{\alpha}_{i,T}}\big]\mathbf{\tilde{s}}_{g},
	\end{aligned}
\end{equation}
where $\bar{\alpha}_{i,T}$ has been defined in \eqref{pre2} and $\mathbf{z}_{0}\in\mathbb{R}^{h\times w\times c}$ is the latent representation encoded from $\mathbf{x}_{0}$, which can be expressed as $\mathbf{z}_{0}=\mathcal{E}(\mathbf{x}_{0})$, where $\mathcal{E}(\cdot)$ is an encoder. In summary, before the transmission beginning, an image is encoded into its latent representation. Then, DSRDM embeds the Gaussian signal into the latent space, performs the forward process and generates $\mathbf{z}_{T}$. Since the embedding process can also follow the same forward statistics as the pre-trained LDM through the proposed training-free algorithm, the receiver can predict the added signal by pre-trained DM directly.

Furthermore, in standard LDMs for image generation, the encoder $\mathcal{E}$ is designed to be powerful and expressive, preserving rich semantic and visual information so that the decoder can reconstruct high-fidelity images. In contrast, the latent $\mathbf{z}_{0}$ is just a carrier in DSRDM. The receiver doesn't need to reconstruct the original image $\mathbf{x}_{0}$, and it only needs to decode the embedded signal. Therefore, we remove the original decoder for image reconstruction of the pre-trained LDM to reduce its decoding latency. By applying the training-free algorithm and latency efficient designs for the proposed DSRDM framework, the entire digital signal transmission process based on DSRDM is shown in Fig. 2. DSRDM not only provides a digital signal transmission approach in SemCom rather than classical image or text transmissions, but also minimizes the training costs and reduces the inference latency of DM, which makes it conveniently to be applied in different scenarios.

\vspace{-2mm}
\section{Simulation Results}

In this section, we evaluate the bits error rate (BER) of the digital signal transmission via DSRDM and its time costs of decoding. To verify the scalability of DSRDM, various communication conditions and pre-trained DMs are considered.

\vspace{-3mm}
\subsection{Simulation Settings}

The carrier images are selected from LSUN Bedroom, ImageNet64 and CIFAR-10 dataset, and the size of their images are $256\times 256$, $64\times 64$ and $32\times 32$, respectively. These datasets are used to trained different pre-trained DMs, and the model weights have been given in \cite{web} for Figs. 4, 5 and 6(b). We also train an LDM on CIFAR-10 for Figs. 3 and 6(a), which model architecture given in \cite{web2}. The pre-trained DMs corresponding to these carrier images of different sizes are applied as encoders and decoders of DSRDM to verify its performance on signal predictions. In the following simulations, to verify the scalability of DSRDM, for the carrier images with the same size, we use the same DM and scheduling strategy without retraining or fine-tuning, regardless of the transmitted digital signal.

Different from the previous DM-based SemCom frameworks which are mainly proposed for image transmission and focused on the image reconstruction quality, DSRDM is design for digital signal transmission and images are only carriers. Thus, BER is the evaluation metric in the simulation. Moreover, to verify the robustness and scalability of DSRDM, in addition to considering different sizes of carrier images and pre-trained DM, we also compare the BER in different communication conditions, such as a) Different modulation schemes (16-QAM \& 256-QAM), and b) different $k$ of Rician fading channel. When $k\to+\infty$, the channel is approximately an AWGN channel, and it becomes Rayleigh channel with $k=0$. Finally,  in BER comparisons, since DSRDM needs match the channel noise and pre-trained DM in its forward process, we select DSRDM without matching channel noise as the benchmark, MAE-based SemCom and DeepIM for binary signal transmission proposed in \cite{mae} and \cite{deepim} as baselines of SemCom and learned decoding approach.
\begin{figure}[!t]
	\vspace{-2mm}
	\centering
	\subfloat[]{\includegraphics[width=1.6in]{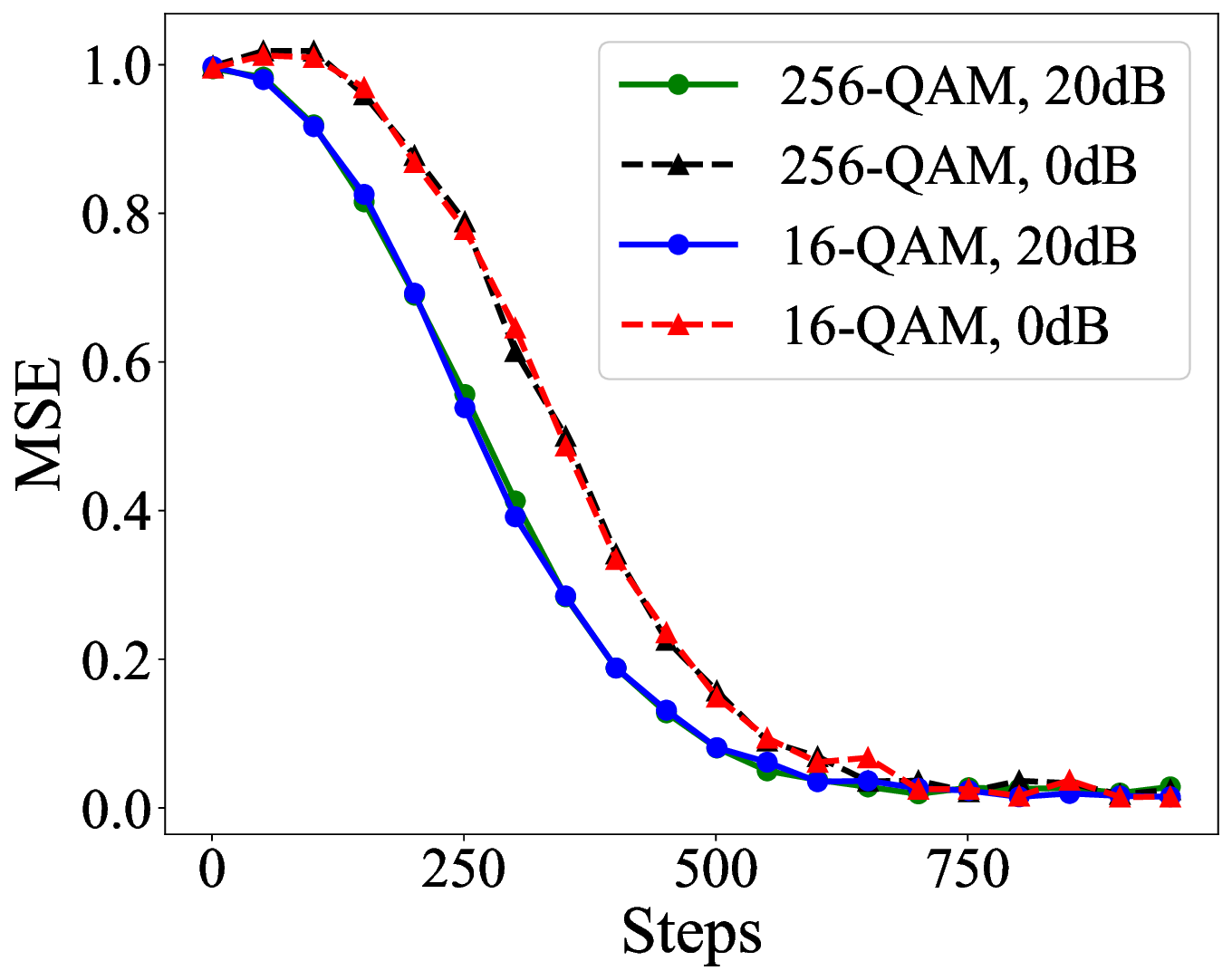}
		\label{fi8_1}}
	\subfloat[]{\includegraphics[width=1.6in]{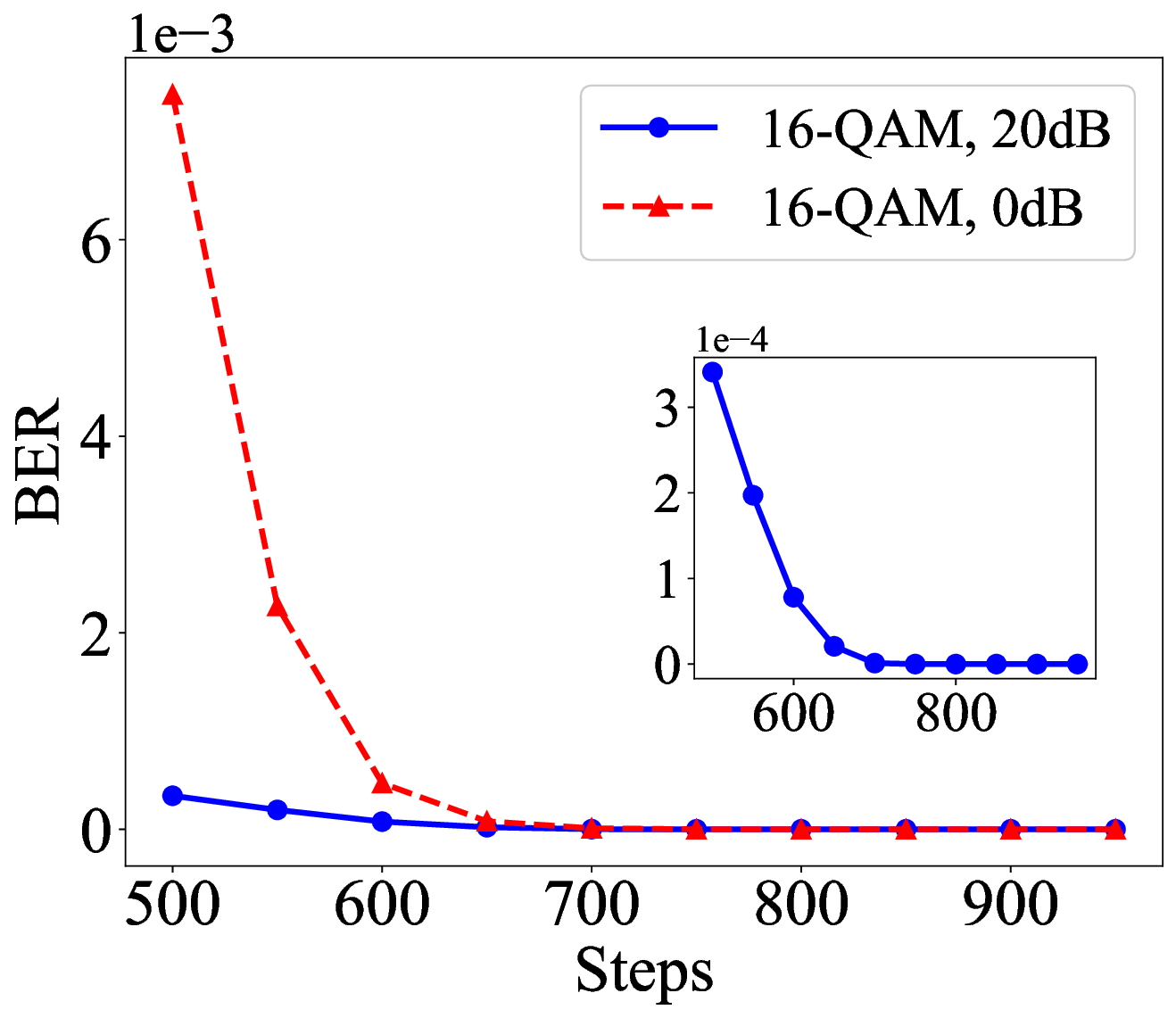}
		\label{fig8_2}}
	\caption{Signal recovery performance of DSRDM over AWGN channels. (a) Prediction MSE of Gaussian signals v.s. inference step on CIFAR-10 dataset. (b) Recovery BER of binary signals v.s. inference step on LSUN dataset.}
\end{figure}
\vspace{-3mm}
\subsection{Simulation Results}

\textit{1) Feasibility analysis:} To verify the feasibility of DSRDM on digital signal transmission, we evaluate the MSE between the predicted Gaussian signal and their original signal on each inference step of DSRDM with different SNRs and Gaussian-modulation schemes over AWGN channels at first. Then, we evaluate their BER after recovering to binary signal from the Gaussian signal. Fig. 3 indicates that digital signals with different Gaussian-modulation schemes (16-QAM \& 256-QAM) can be well-predicted and recovered by DSRDM with the same pre-trained DM and scheduling strategy. From Fig. 3(a) we know that, not all the inference steps during the inverse process of DM could achieve an outstanding performance on Gaussian signal prediction, especially at the beginning of the inverse process, but the prediction MSE could achieve low values (less than 0.15) after around 500 steps. Thus, we could add the Gaussian signal at the last 500 noise-adding steps of DM for transmission. The BER on all inference steps from 500 to 1000 are given by Fig. 3(b). The value of BER of each step decreasing quickly after 500 inference steps. Although there is still a gap between $20$dB and $0$dB SNR, ($7.75\times 10^{-3}$ and $3.51\times 10^{-4}$), they decrease to less than $1{\times} 10^{-5}$ rapidly. This result indicates that adding Gaussian signals at the first 500 steps of DM and recover them at the receiver is feasible and performs well. Thus, in following simulations, we also use the first 500 steps for Gaussian signal adding. Specifically, after the raw bits modulated to Gaussian signals, they are only added to carrier images on the first 500 forward steps, and the rest 500 steps are adding random noise.  Similarly, only the predicted Gaussian signals from the last 500 steps on the reverse process are used to recover raw bits.

\begin{figure}[!t]
	\vspace{-3mm}
	\centering
	\subfloat[]{\includegraphics[width=1.6in]{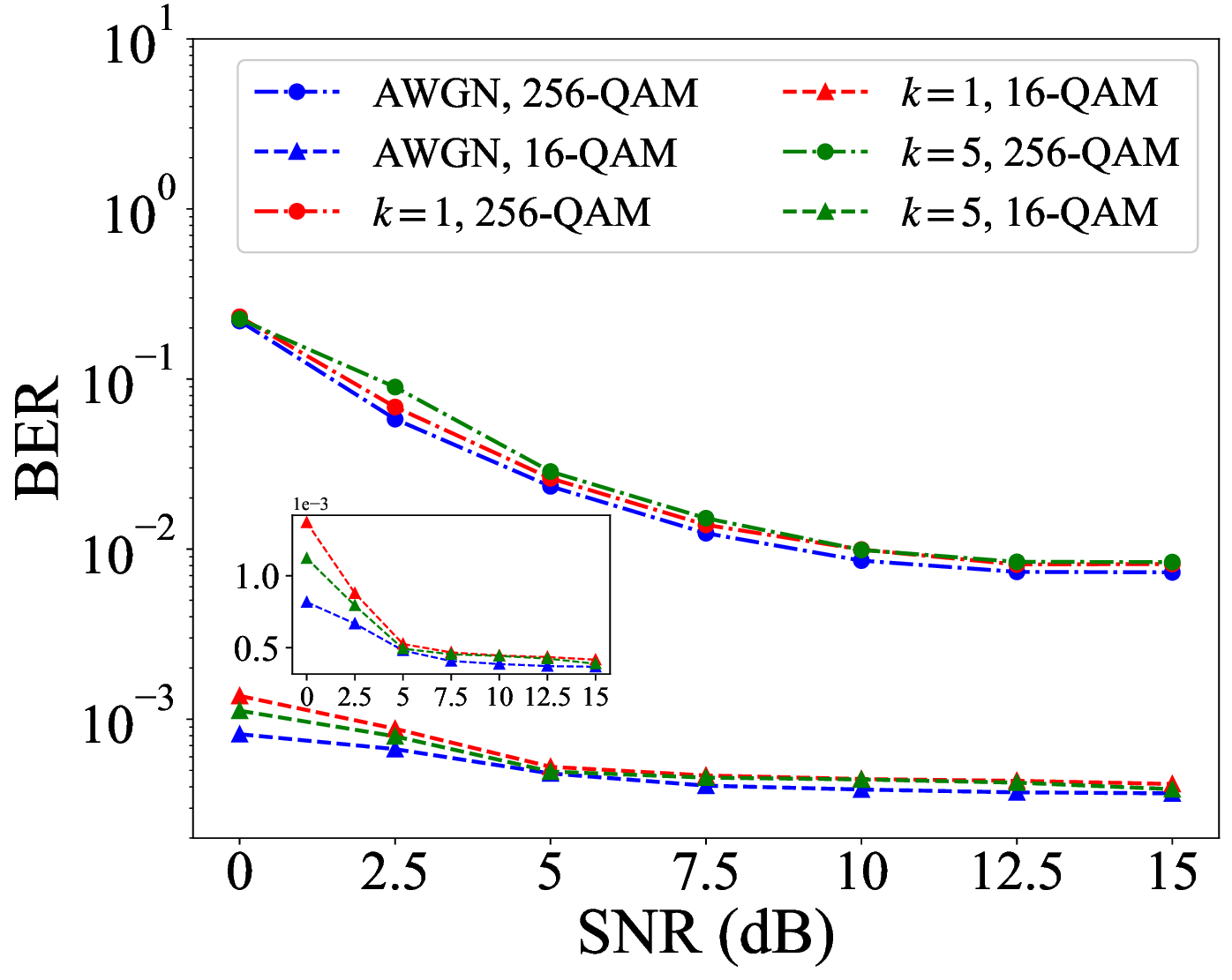}
		\label{fi8_3}}
	\subfloat[]{\includegraphics[width=1.6in]{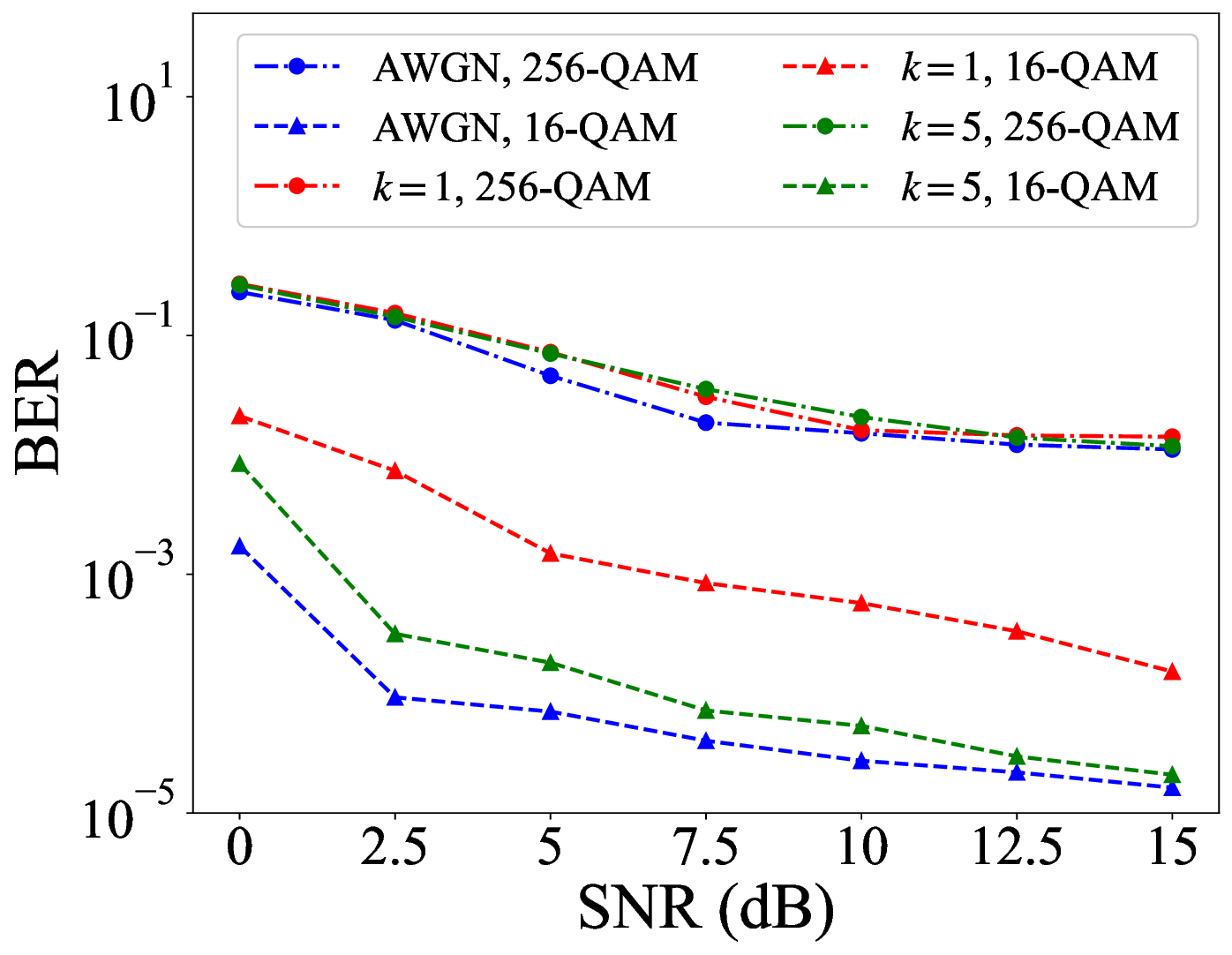}
		\label{fig8_4}}
	\caption{ Recovery BER of binary signals v.s. SNR. (a) Recovery BER on CIFAR-10 dataset. (b) Recovery BER on ImageNet64 dataset.}
\end{figure}

\textit{2) Performance evaluations:} Fig. 4 shows the BER of proposed DSRDM with various SNR on different Gaussian-modulation schemes, (16-QAM \& 256-QAM), 
channel conditions (AWGN \& Rician fading channels) 
and size of carrier images. ($32\times 32$ \& $64\times 64$). 
In Fig. 4, DSRDM with 16-QAM over the AWGN channel consistently achieves the lowest BER compared to other cases. From the perspective of the modulation scheme, this is because 256-QAM is more sensitive than 16-QAM to channel noise, since each point has a smaller Euclidean distance to neighbors with the same transmit power. Thus, although the recovered MSEs of the prediction Gaussian signal of 16\&256-QAM by DSRDM are similar, 256-QAM takes less noise power to cause a decision error, which results in a higher BER. 

\textit{3) Time costs evaluations:} In Fig. 5(a), we compare the inference time costs of DM-based and LDM-based DSRDM to verify the effectiveness of the training-free algorithm on reducing inference complexity. The comparison results show that LDM-based DSRDM has faster inference speed. In addition, since the latency of diffusion inference is limited by the model architecture, hardware design and sampling strategy. Thus, DSRDM is more suitable for some tasks without real-time requirements, such as the parameters/gradients transmission between users and base station in federated learning \cite{fl}.

\textit{4) Performance comparisons:} Finally, to verify the performance of the proposed DSRDM the training-free algorithm on digital signal transmission and pre-trained DM direct application, we compare the BER between DSRDM and baselines in Fig. 5(b). The comparison results show that, DSRDM with channel noise matching has consistently lower BER, even though all compared schemes perform well at higher SNRs.
\section{Conclusions}

In this letter, we propose and investigate DSRDM for digital signal transmission in SemCom systems. By replacing the added noise in the diffusion process with Gaussian signals, DSRDM enables effective encoding and recovery of general digital signals through forward and reverse diffusion. To further reduce complexity, a signal adding approach and an LDM-based scheme are introduced. Simulation results verify the feasibility and good performance of applying diffusion models to general semantic signal transmission.

\vspace{-2mm}

\begin{figure}[!t]
	\vspace{-3mm}
	\centering
	\subfloat[]{\includegraphics[width=1.3in]{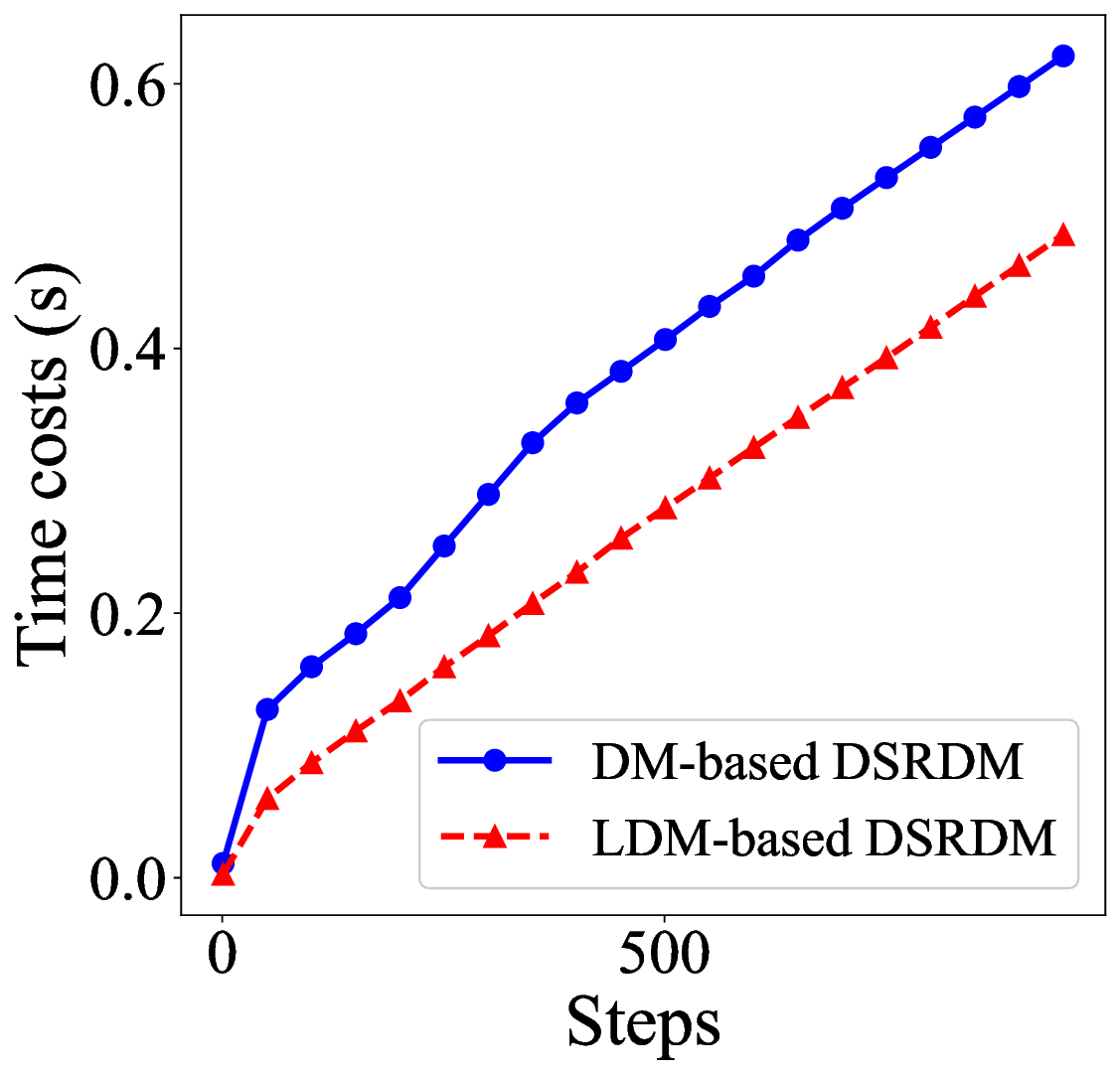}
		\label{fi8_5}}
	\subfloat[]{\includegraphics[width=2.2in]{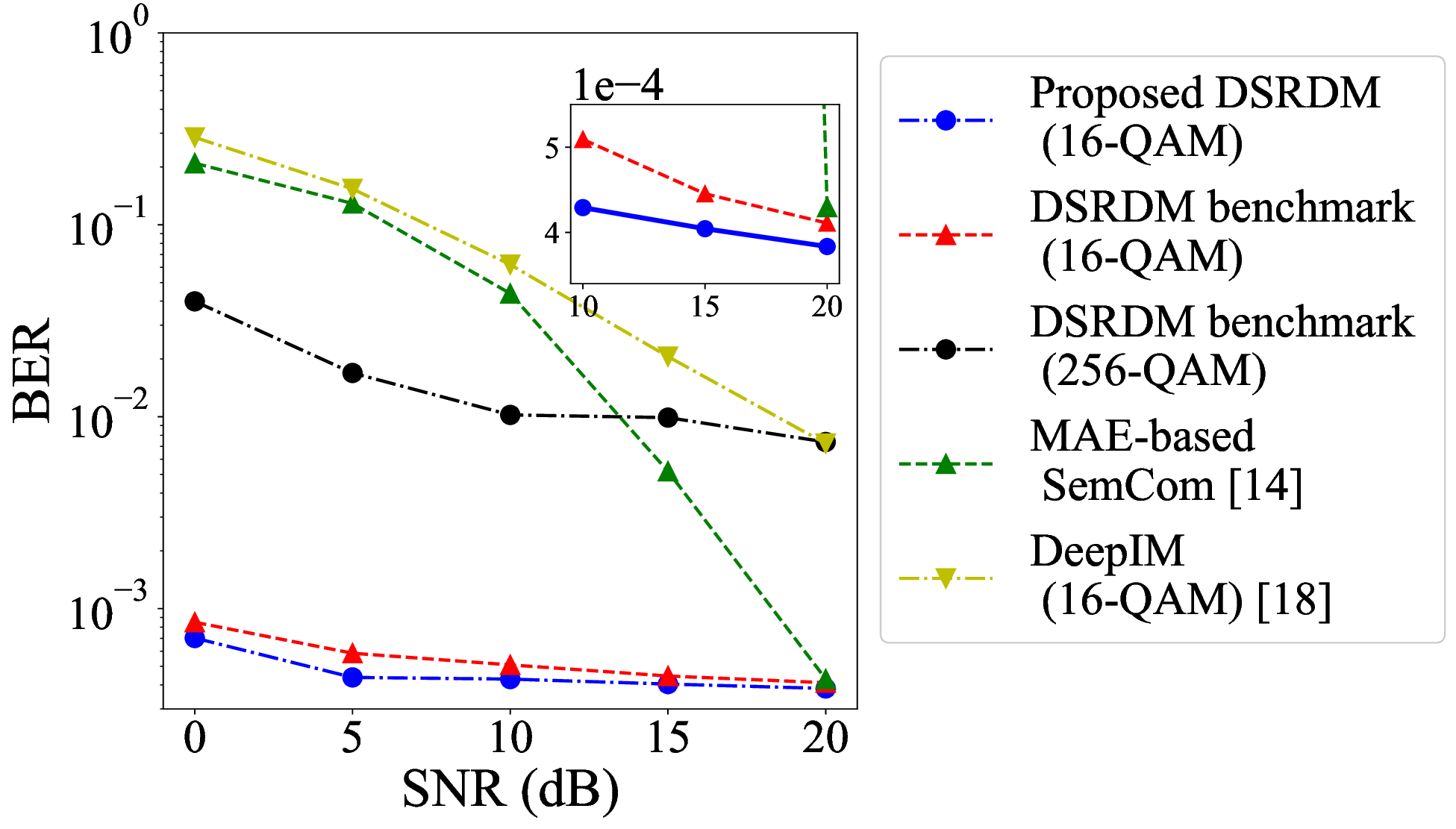}
		\label{fig8_6}}
	\caption{Inference time and BER comparison over AWGN on CIFAR-10 dataset. (a) Inference time v.s. inference step. (b) Recovery BER v.s. SNR.}
\end{figure}

\bibliographystyle{ieeetr}
\bibliography{ref2}

@ARTICLE{diffusion,
  author={Liang, Chengyang and others},
  journal={IEEE Commun. Surv. Tut.}, 
  title={Generative AI-Enabled Semantic Communication: State-of-the-Art, Applications, and the Way Ahead}, 
  year={2026},
  volume={28},
  number={},
  pages={3976-4015},
  doi={10.1109/COMST.2025.3649707}}

@ARTICLE{image1,
	author={Liang, Chengyang and others},
	journal={IEEE Commun. Lett.}, 
	title={Selection-Based Image Generation for Semantic Communication Systems}, 
	year={2024},
	volume={28},
	number={1},
	pages={34-38},
	doi={10.1109/LCOMM.2023.3339534}}

@ARTICLE{image2,
  author={Guo, Lei and others},
  journal={IEEE Trans. Wireless Commun.}, 
  title={Diffusion-Driven Semantic Communication for Generative Models With Bandwidth Constraints}, 
  year={2025},
  volume={24},
  number={8},
  pages={6490-6503},
  doi={10.1109/TWC.2025.3553851}}

@INPROCEEDINGS{audio,
  author={Grassucci and others},
  booktitle={Proc. IEEE International Conference on Acoustics, Speech and Signal Processing (ICASSP)}, 
  title={Diffusion Models for Audio Semantic Communication}, 
  year={2024},
  volume={},
  number={},
  pages={13136-13140},
  doi={10.1109/ICASSP48485.2024.10447612}}

@ARTICLE{video,
  author={Li, Nan and others},
  journal={IEEE Trans. Wireless Commun.}, 
  title={Goal-Oriented Semantic Communication for Wireless Video Transmission via Generative {AI}}, 
  year={2026},
  volume={25},
  number={},
  pages={10841-10854},
  doi={10.1109/TWC.2026.3654401}}

@ARTICLE{text,
	author={Xie, Huiqiang and others},
	journal={IEEE Trans. Signal Process.}, 
	title={Deep Learning Enabled Semantic Communication Systems}, 
	year={2021},
	volume={69},
	number={},
	pages={2663-2675},
	doi={10.1109/TSP.2021.3071210}}

@ARTICLE{dm,
  author={Liang, Chengyang and others},
  journal={IEEE Trans. Commun.}, 
  title={Image Generation With Supervised Selection Based on Multimodal Features for Semantic Communications}, 
  year={2025},
  volume={73},
  number={12},
  pages={14469-14485},
  doi={10.1109/TCOMM.2025.3615798}}

@INPROCEEDINGS{dl,
  author={Dara, Suresh and others},
  booktitle={Proc. International Conference on Electronics, Communication and Aerospace Technology (ICECA)}, 
  title={Feature Extraction By Using Deep Learning: A Survey}, 
  year={2018},
  volume={},
  number={},
  pages={1795-1801},
  doi={10.1109/ICECA.2018.8474912}}

@ARTICLE{semcom,
	author={Yu, Xianhua and others},
	journal={IEEE Internet Things J.}, 
	title={A Novel Lightweight Joint Source-Channel Coding Design in Semantic Communications}, 
	year={2025},
	volume={12},
	number={11},
	pages={18447-18450},
	doi={10.1109/JIOT.2025.3556909}}

@ARTICLE{cddm,
	author={Wu, Tong and others},
	journal={IEEE Trans. Wireless Commun.}, 
	title={{CDDM}: Channel Denoising Diffusion Models for Wireless Semantic Communications}, 
	year={2024},
	volume={23},
	number={9},
	pages={11168-11183},
	doi={10.1109/TWC.2024.3379244}}

@INPROCEEDINGS{scdm,
  author={Mo, Hao and others},
  booktitle={Proc. IEEE International Conference on Communications (ICC)}, 
  title={{SCDM: S}core-Based Channel Denoising Model for Digital Semantic Communications}, 
  year={2025},
  volume={},
  number={},
  pages={3772-3778},
  doi={10.1109/ICC52391.2025.11161428}}

@INPROCEEDINGS{dmce,
	author={Zeng, Youcheng and others},
	booktitle={Proc. IEEE International Conference on Communications (ICC)}, 
	title={{DMCE}: Diffusion Model Channel Enhancer for Multi-User Semantic Communication Systems}, 
	year={2024},
	volume={},
	number={},
	pages={855-860},
	doi={10.1109/ICC51166.2024.10622730}}

@ARTICLE{fl,
	author={Yan, Zhigang and others},
	journal={IEEE Trans. Commun.}, 
	title={Performance Analysis for Resource Constrained Decentralized Federated Learning Over Wireless Networks}, 
  	year={2024},
	volume={72},
	number={7},
	pages={4084-4100},
	doi={10.1109/TCOMM.2024.3362143}}

@ARTICLE{mae,
	author={Yan, Zhigang and others},
	journal={IEEE Wireless Commun. Lett.}, 
	title={Semantic Communications for Digital Signals via Carrier Images}, 
	year={2025},
	volume={14},
	number={6},
	pages={1816-1820},
	doi={10.1109/LWC.2025.3557843}}

@INPROCEEDINGS{msm,
  author={Ye, Weitan and others},
  booktitle={Proc. IEEE International Conference on Communications Workshops (ICC Workshops)}, 
  title={Model Shift Modulation for Semantic Communication}, 
  year={2025},
  volume={},
  number={},
  pages={1717-1722},
  doi={10.1109/ICCWorkshops67674.2025.11162462}}

@misc{web,
  author       = {Bao, Fan},
  title        = {{Extended-Analytic-DPM}},
  howpublished = {GitHub repository: https://github.com/baofff/Extended-Analytic-DPM},
  year         = {2022},
  url          = {{https://github.com/baofff/Extended-Analytic-DPM}},
}

@misc{web2,
  author       = {Chatta, Nikhil},
  title        = {{LatentDiffusion}},
  howpublished = {GitHub repository: https://github.com/nikhilchatta/LatentDiffusion},
  year         = {2025},
  url          = {{https://github.com/nikhilchatta/LatentDiffusion}},
}

@ARTICLE{deepim,
  author={Luong, Thien Van and others},
  journal={IEEE Wireless Commun. Lett.}, 
  title={Deep Learning-Based Detector for {OFDM-IM}}, 
  year={2019},
  volume={8},
  number={4},
  pages={1159-1162},
  doi={10.1109/LWC.2019.2909893}}

\vfill

\end{document}